\documentclass{llncs}
\usepackage{color}
\usepackage{epsfig}
\usepackage{graphics}

\begin{document}
\title{FreePub: Collecting and Organizing Scientific Material Using Mindmaps
\\ {\footnotesize(or ...where mindmaps meet web search)}}
%
%
\author{Theodore Dalamagas\inst{1} \and Tryfon Farmakakis\inst{2} \and 
Manolis Maragkakis\inst{3} and Artemis G. Hatzigeorgiou\inst{3}}
%
%
%
\institute{IMIS Institute/``Athena'' R.C., Athens, Greece\\
\email{dalamag@imis.athena-innovation.gr}
\and
University of Edinburgh, School of Informatics, Edinburgh, Scotland \\
\email{T.Farmakakis@sms.ed.ac.uk}
\and
BSRC Fleming, Vari, GR \\
\email{(maragkakis,artemis)@fleming.gr}
}

\maketitle              

\begin{abstract}
This paper presents a creativity support tool, called \texttt{FreePub}, to collect and organize
scientific material using mindmaps. Mindmaps are visual, graph-based represenations of concepts,
ideas, notes, tasks, etc. They generally take a hierarchical or tree branch format, with ideas
branching into their subsections. \texttt{FreePub} supports creativity cycles. A user starts such a cycle by setting up her domain of interest using mindmaps. 
Then, she can browse mindmaps and launch search tasks to gather relevant publications from several data sources. \texttt{FreePub}, besides publications,  identifies helpful supporting material (e.g., blog posts, presentations). All retrieved information
from \texttt{FreePub} can be imported and organized in mindmaps. \texttt{FreePub} has been fully implemented on top
of FreeMind, a popular open-source, mindmapping tool.
\end{abstract}
\section{Introduction}

Web search engines are widely used for searching information on the Web. Their increased popularity is due to the following reasons: the search model employed (i.e., keyword-based) is simple and 
easy to use, and the search techniques are nowadays mature enough to support fast text retrieval with accurate results. 

However, there are use cases where the information need is complex. Consider, for instance,
a researcher that needs to set up her research agenda and generate innovative ideas. She often has 
the ``big picture'' of the domain, i.e., an abstraction based on topics, thoughts, and 
everything else that helps setting up her search plan to explore the domain. Based on this initial 
abstraction, she (a) gathers information from several data sources, (b) organizes the information, 
(c) generates hypothesis and scientific results, (c) disseminates those results, and then
(d) starts over by refining her abstraction and search plan. Such a \emph{creativity cycle} actually 
enables discovery and innovation. 

To illustrate an example of a creativity cycle, consider a researcher interested in \emph{sequence matching techniques} for genomics, and the following use case: 
\begin{enumerate}
\item The researcher starts by looking for journal papers that make a thorough review of this particular research area (i.e., the so-called survey papers), and blog articles that provide a review of the current state-of-the-art technologies technologies. 
\item After organizing and studying the retrieved material, she pays more attention to the
\emph{local alignment} problem, that is ``given a query sequence and a data sequence, find pairs of 
similar subsequences chosen from these sequences''. She finds out that the dynamic programming
solutions suggested to deal with that problem have high computational cost, and that this is the
reason for researchers to work on approximation solutions (i.e., methods to return some but not all
of the alignment results, according to some statistical significance model).
Thus, she starts now looking for papers related to \emph{approximate local alignment}.
\item After organizing and studying the retrieved material, she concludes that those methods, athough efficient, are not appropriate for several cases where the full result set of alignments is needed.
Thus, she starts now looking for papers that are related to \emph{indexing schemes for efficient local alinment}. These approaches exploit data structures which speed up the matching process
between a large data sequences and a query sequence, at the expense of having to maintain these structures when data changes.
\item At any step of the above creativity cycle, she disseminate her findings to other researchers to get feedback.
\end{enumerate}

New search models and techniques are necessary to support creativity and innovation \cite{shneiderman-acmcomm07}. A critical objective is to support creativity cycles, and also to provide effective presentation and visualization capabilities for the lists of retrieved resources that will guide users during their search and exploration.  

{\em Mindmapping} \cite{buzan96,farrand-me02} makes use of visual diagrams to capture and organize information. They generally take a hierarchical or tree branch format, with ideas branching into their subsections. Mindmapping elements include concepts, ideas, notes, tasks, etc. One can use mindmaps
to summarizing information, consolidating information from different research sources, thinking
through complex problems, and presenting information showing the overall structure of her topic.
Mindmaps is an excellent model for visualize, structure, and classify ideas, and support
creative thinking.

This paper presents a creativity support tool, called \texttt{FreePub}, to collect and organize
scientific material using mindmaps. \texttt{FreePub} supports creativity cycles, assisting users to:
\begin{enumerate}
\item set up their domain of interest using mindmaps, 
\item browse mindmaps and launch search tasks to gather  relevant documents from several data 
sources, 
\item identify supporting material for those documents (e.g., blog posts, presentations), and 
\item import and organise all retrieved information in mindmaps.  
\end{enumerate}

\texttt{FreePub} (http://web.imis.athena-innovation.gr/projects/mm/) has been built on top of Freemind \cite{fm}, a popular open-source, mindmapping tool.

\noindent\textbf{Outline.} In the next section we give an overview of \texttt{FreePub}
architecture, and we discuss the related work.
Section \ref{sec:mm} describes mindmaps. Section \ref{sec:search} presents the search facilities of
\texttt{FreePub}, and Section \ref{sec:sqe} describes the semantic query expansion mechanism.
Section \ref{sec:freepub} discusses a test case for \texttt{FreePub}, and, finally, Section \ref{sec:con}
concludes the work.

\section{Overview and Related Work}
In this section we give a brief overview of tool features and technologies used, and we discuss
the related work. 

\begin{figure}[!ht]
{\center
\includegraphics[width=9cm]{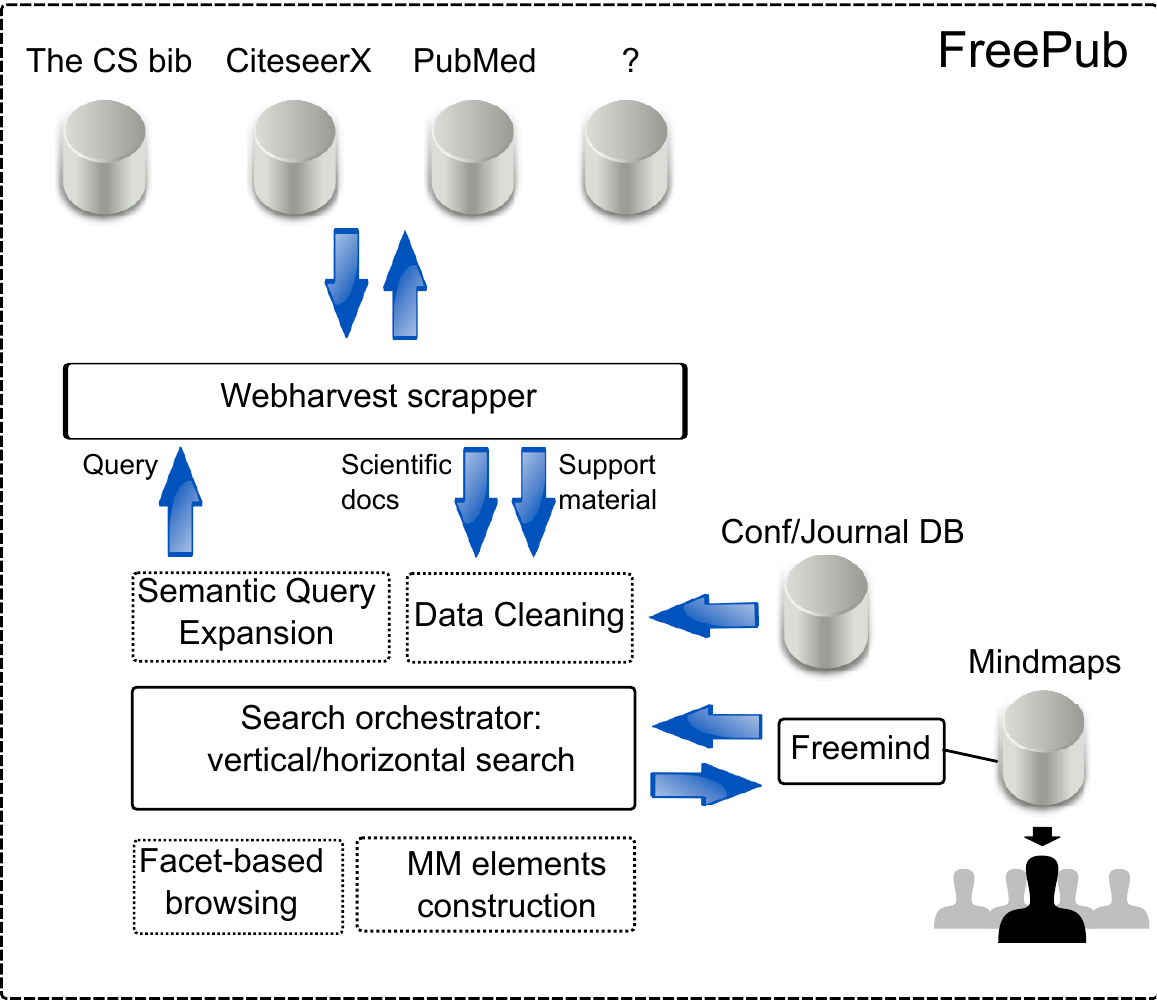}
\caption{\texttt{FreePub}'s architecture.}
\label{fig:architecture}
}
\end{figure}

Figure \ref{fig:architecture} shows the architectute of \texttt{FreePub}.
\texttt{FreePub} has been implemented on top of FreeMind~\cite{fm}.
Freemind provides an excellent user-friendly editor to build mindmaps. Users exploit mindmaps to set up their knwoledge domain, and collect and organize
scientific material retrieved from several data sources. The \emph{search orchestrator module} is 
responsible for launching \emph{vertical} and \emph{horizontal} search tasks, and coordinate their 
operation in order
to retrieve publications and supporting material. The \emph{semantic query expansion module} 
provides
intelligent retrieval facilities by enriching user queries with terms extracted from mindmap 
elements
to improve search effectiveness. The \emph{data cleaning module} processes the result lists to 
remove 
name ambiguities and inconsistencies, and also to remove duplicate results. \texttt{FreePub}
maintains a database of conference/journal info to assist cleaning tasks.
The \emph{facet-based browsing module} provides visualization options using several information 
facets to present the results. Finally, the \emph{MM element construction module} is responsible for 
transfering the result lists into the mindmaps, according to user needs.

The use of mindmaps in information retrieval tasks has been acknowledged by several researchers.
In~\cite{beel-col09}, the authors present how information retrieval on mind maps could be used
to enhance expert search, document summarization, keyword based search engines, document
recommender systems and determining word relatedness.

Also,~\cite{bia-ecdl10} describes how one can use mindmaps 
to succesfully model, design, modify, import and export XML DTDs, XML schemas and XML
dooc, getting very manageable, easily comprehensible, folding diagrams. They actually converted
a general purpose mind-mapping tool into a very powerful tool for XML vocabulary design and
simplification. Finally, SciPlore MindMapping~\cite{beel-dlib09} is the first mind mapping tool
focusing on researchers’ needs by integrating mind mapping with reference and pdf management.
SciPlore MindMapping offers all the features one would expect from a standard mind mapping
software, plus the following special features for researchers: adding reference keys, PDF
bookmark import, and monitoring folders for new pdfs.

Compared to the above works, \texttt{FreePub} provides a full-fledged retrieval service to collect
scientific material using mindmaps. It retrieves not only relevant publications, but also supporting 
material, like blog posts, presentation slides, from several wrapped data sources. Also, it exploits 
a semantic query expansion mechanism to enrich user queries with mindmap element terms for improved search effectiveness. 

There are also several open source (e.g., Vue, 
XMind, Compendium\footnote{http://vue.tufts.edu/,
http://www.xmind.net/, http://compendium.open.ac.uk/})
and commercial tools (e.g., MindManager, ConceptDraw,
iMIndMap\footnote{http://www.mindmanager.com,
http://www.conceptdraw.com,\\
http://www.thinkbuzan.com
}) for mindmapping. However, they are 
actually mindmapping editors, providing advanced visualization capabilities, document handling and 
integration facilities with other popular software suites. Neither of them exploits mindmaps as a means for 
exploration Web search, giving also intelligent query expansion mechanisms, like FreePub does. 

\section{Mindmapping}
\label{sec:mm}

{\em Mindmapping}  \cite{buzan96,farrand-me02} refers to graphical representations of 
\emph{elements} such as concepts, ideas, notes, tasks, or other items related to a topic of study. 
Mindmapping elements are organized in hierarchical branches or groups according to
the semantic interpretation given by the user. However, everything is built around a central topic 
or idea.
The key feature of mindmapping is that the elements are arranged in a non-linear fashion.
Thus, users are free to enumerate and connect concepts without a tendency to begin within a 
particular conceptual framework. This encourages a brainstorming approach to planning and 
organizational tasks, and idea generation. 

Mindmaps is an excellent model for setting up workspaces for internet search, project and task 
management (including links to necessary files, executables, source of information), knowledge base 
organization (notes, references), and essay writing and brainstorming. They allow for greater 
creativity when recording ideas and information, and help the note-takers to associate topics and 
ideas with visual representations. 

A key difference between mindmaps and other graph-based formal modelling representations, e.g. UML, 
semantic networks, TopicMaps, is that the the latter have explicit structured elements to model
relationships. Contrary, mindmaps rerpesent the \emph{visual mnemonics} of users, exploiting colors, 
icons and informal visual representations.
Visual methods like mindmaps have been used for centuries in learning and problem solving
by educators for recording knowledge, visual thinking, and problem solving. 
Also, mindmaps are based on radial hierarchies showing connections with a centered ruling concept.

\begin{figure}[!ht]
{\center
\fbox{\includegraphics[width=12cm]{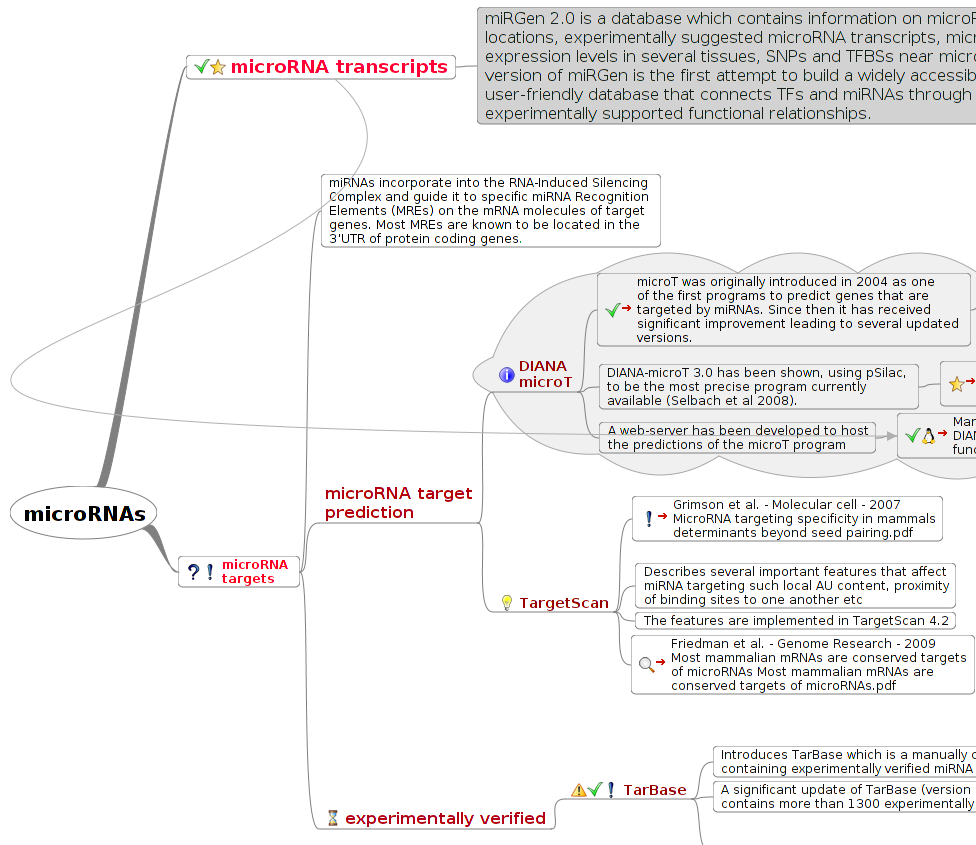}}
\caption{A mindmap example.}
\label{fig:mindmap-example}
}
\end{figure}

Freemind~\cite{fm} provides a user-friendly editor to build mindmaps. 
Table \ref{tab:10} presents the most important mindmap elements used by Freemind. Figure \ref{fig:mindmap-example} shows a mindmap example, organizing information about
microRNA entities (see also Section \ref{sec:freepub}). In this mindmap, for example,
\texttt{microRNA} is the central idea where all other elements are structured around. 
\texttt{microRNA targets} and \texttt{microRNA transcripts} are topic elements, while 
\texttt{microRNA target prediction} is a sub\-topic element. The text ``miRNA incorporate into the 
RNA-Induced...'' is a detail element.
\begin{table}[!ht]
\begin{center}
{\scriptsize
\begin{tabular}{|p{4cm}|p{4cm}|}
\hline \textbf{Topic, Larger topic}: main elements, arranged in a topic/subtopic fashion, to represent ideas & \textbf{Waiting topic}: a topic that needs to be reconsidered   \\
\hline \textbf{Needs action}: an element for which action is needed & \textbf{Hot}: a critical element   \\
\hline \textbf{Detail}: text content element (e.g., notes, abstracts, etc) & \textbf{Link}: direct link element to user folders, urls, or local files  \\
\hline \textbf{Object (keywords)}: set of words used as keywords for an element & \textbf{Object (code)}: piece of code for an element   \\
\hline \textbf{Question}: issues that need to be considered for an element & \textbf{Cloud}: set of related elements    \\
\hline
\end{tabular}
}	
\end{center}
\caption{Mindmap elements in FreeMind.}
\label{tab:10}
\end{table}%

\section{Searching facilities}
\label{sec:search}

As the user explores a mindmap, she can initiate a search
task to retrieve, from several wrapped data sources, documents relevant to mindmap topics. Various search parameters can be determined, like the number of results, the 
data sources used, etc. For each search task, \texttt{FreePub} starts
the retrieval service by first formulating the necessary queries.
Keywords are extracted from the content of mindmap elements selected by the user in order to form keyword queries to send to the data sources. A key feature of \texttt{FreePub} is a semantic query expansion mechanism
used to extract keywords not only from the selected mindmap elements, but also
from their \emph{semantic neighbourhood}. We discuss this feature in detail later on,
in Section \ref{sec:sqe}.

\noindent\textbf{Vertical search}. Keyword queries are sent to all wrapped
data sources to retrieve relevant documents. Such data sources usually
provide \emph{vertical search facilities}, i.e., tailored to certain types
of information resources - in our case, computer science publications (e.g.,  
DBLP, PubMed~\cite{dblp,pubmed}). 
\texttt{FreePub}
wraps data sources using WebHarvest~\cite{wh}. We discuss wrapping facilities later on.

The resulting snippets are extracted from the data sources, cleaned, and presented to the user. 
Cleaning includes several facilities used to process the results in order to remove ambiguities, 
inconsistencies, etc. Specifically, the system utilizes a catalog with journal names and conferences 
extracted from DBLP and PubMed~\cite{dblp,pubmed} to deal with name inconsistencies. Each journal/conference 
name in the snippets is matched again this catalog to determine a common name for all snippets. The 
catalog actually maintains two string values for each journal/catalog entry: a short string for the 
acronym and a long one for the title of the entry. 

Matching is based on the Levenshtein distance~\cite{gusfield1999} $L$ between two strings.
The Levenshtein distance is defined as the minimum number of edit operations needed to transform one string into the other, with the allowable edit operations being insertion, deletion, or substitution of a single character. For example, $L$(``VLDD'', ``VLDB Conf'')$=6$: replace
`D' with `B', and insert ` ', `C', `o' `n' `f', a total number of $6$ operations.   

Assuming a string $s$ and a catalog 
of $n$ entries $\{(a_1,t_1), (a_2,t_2), \ldots, (a_n,t_n)\}$ with pairs of acronyms $a_i$ and 
titles $t_i$, $s$ is matched to the entry $(a_i,t_i)$ such that $L(s,a_i)+L(s,t_i)$ is minimized 
($0<i\leq n$). For example, ``Very Large Database Conf'' and ``VLDB Conf'', both are matched to 
(``Very Large Da\-ta\-ba\-se Conference'', ``VLDB'') catalog entry.

\noindent\textbf{Duplicate elimination}. 
Since results are retrieved from several data sources, duplicate results may appear.
Duplicates are removed using entity resolution blocking techniques~\cite{euijong-sigmod09}.
The problem of entity resolution involves finding records in a dataset
that represent the same real-world entity. Blocking techniques
divide data into groups and only compares records within the same group, to avoid redundant
comparisons. This is based on the assumption that records in different blocks are unlikely to match. 

FreePub implements the following efficient strategy for entity identification and duplicate elimination:
\begin{enumerate}
\item The result list of each data source is partitioned into groups, using the publication
date as key for each group. For each group we maintain a \texttt{(key$\rightarrow$value)} hash structure $H$, where \texttt{key} is the date and \texttt{value} is the list of publication
objects $o_i$. For example: $H_1=(2004\rightarrow \{o_1, o_3, o_5, o_6\})$, $H_2=(2005\rightarrow 
\{o_2, o_4\})$ for data source $1$, $H_3=(2004\rightarrow \{o_1, o_5, o_8\})$ for data source $2$, 
etc.
\item Then, to identify duplicates we check pairs of publication objects $(o_i, o_j)$ only for objects than share the same key (date). Checking is done using exact string matching
on publication title and publication forum. For instance, in the previous example, 
only pairs of publication objects from $H_1$ value list and $H_3$ value list will be checked. 
\end{enumerate}
\noindent\textbf{Horizontal search}. After retrieving docuements relevant to mindmap elements, the user may launch another search task to get supporting material for these
documents. Such material includes blog posts discussing the topic of a document, related presentations, other reports etc. To detect the material, \texttt{FreePub}
uses \emph{horizontal search facilities}, i.e., general search engines that cover all the Web,
and appropriate options to restrict searches to only certain type of documents.
Specifically, FreePub searches for the following support material for each retrieved
publication:

\begin{enumerate}
\item pub document: a query string is constructed from publication's title, and the 
filetype:pdf or doc option is used in order to retrieve results.
Further heuristic rules are used in order to certify that the retrieved result is indeed the
document of the publication. E.g., we parse the retrieved documents and check whether the title of the publication appears in, etc.
\item pub abstract: the abstract is extracted either by parsing the document identified in 1. or by looking for the appropriate metadata fields in the data source used, since
several data sources provide such information. 
\item slide presentation:  a query string is constructed from publication's title, and 
the filetype:ppt or pdf option is used in order to retrieve results. Further heuristic
rules are used in order to certify that the retrieved results are indeed presentations.
E.g., we parse the retrieved documents and check whether certain terms appear inside, e.g.,
the term``outline'', terms from the sections of the document identified in 1., etc.
\item blog entries: a query string is constructed from publication's title along
with author's name and issued to the Google Blogs Search Engine to retrieve results.
\end{enumerate}

\noindent\textbf{Wrappers}.
\texttt{FreePub} retrieves scientific documents from several data sources,
e.g., the collection of Computer Science Bibliography~\cite{csbib}, 
citeseerX~\cite{citeseerx}, and PubMed~\cite{pubmed}. New data sources can be easily integrated.
\texttt{FreePub} wraps data sources using WebHarvest~\cite{wh}, a Web scraping tool that (a) captures data source search capabilities, and (b) simplifies Web information extraction from data sources. WebHarvest provides several types of processors (e.g., html-to-xml, xpath, etc) to define a sequence of extraction operations on Web pages and
identify the required html parts easily.

To demonstrate how WebHarvest work, we show the part of the html source of the first three
results returned from google blog search for the term ``ubuntu''.  

{\scriptsize
\begin{verbatim}
...1st result
<a href="http://www.howtoforge.com/how-to-upgrade-ubuntu-10.04-..." id="p-1">
How To Upgrade <b>Ubuntu</b> 10.04 (Lucid Lynx) To 10.10 (Maverick Meerkat)
(Desktop; Server)<br>
</font>
<font size=-1>
...2nd result
<a class=f1 href="http://www.howtoforge.com/" id="pb-1"
title="http://www.howtoforge.com/">
HowtoForge - Linux Howtos and Tutorials - -
http://www.howtoforge.com/</a>
</font>
</td>
</tr>
</table>
<p class=g></p>
...3rd result
<a href="http://www.readwriteweb.com/cloud/2010/10/latest-ubuntu-1010-emphasizes.php"
id="p-2">
Latest <b>Ubuntu</b> 10.10 Emphasizes the Cloud - ReadWriteCloud</a>
<table border=0 cellpadding=0 cellspacing=0><tr><td class=j>
<font color=#555555 size=-1>11 hours ago </font>
<font color=#555555 size=-1>by Audrey Watters</font><br><font size=-1>
Open source operating system <b>Ubuntu</b> 10.10 is available to download today for desktop,
notebook, and server editions. Hooray for well-timed 10.10;s. All these versions are
emphasizing Canonical embracing
... 
\end{verbatim}
}

WebHarvest is based on an XML configuration file describing the process to extract data. The elements define access to html pages, files, databases, mails, ftp servers and configures the work flow. An example of an XML configuration file that parses the above html source follows:

{\scriptsize
\begin{verbatim}
1.  <?xml version="1.0" encoding="UTF-8"?> 
2.  <config charset="UTF-8"> 
3.  <var-def name="searchQuery" overwrite="false"/> 
4.  <var-def name="content"> 
5.  <html-to-xml> 
6.   <http url="http://blogsearch.google.com/blogsearch?hl=
        en&amp;oi=spell&amp;ie=UTF-8&amp;q=\${searchQuery}&amp;btnG=Search+Blogs"/> 
7.  </html-to-xml> 
8.  </var-def> 
9. <var-def name="results1"> 
10.  <xpath expression="//a[contains(@id,'p-')]"> 
11.   <var name="content"/> 
12.  </xpath> 
13. </var-def> 
14. <var-def name="results2"> 
15.  <xpath expression="//td[@class='j']"> 
16.   <var name="content"/> 
17.  </xpath> 
18. </var-def> 
19. </config>
\end{verbatim}
}

In line 3, the variable \texttt{searchQuery} is assigned the value ``ubuntu'', which is actually
the search term. In line 6, the value of \texttt{searchQuery} is appended to the Google blogs search 
address and passed to the WebHarvest’s HTTP engine which returns the results page in raw HTML. In 
line 5, WebHarvest’s HTML-to-XML engine is called, which transforms the raw HTML code into a well 
formed XML document, which is assigned to the newly defined variable \texttt{content} in line 4. An 
abstract of the XML document that contains the information for one result is shown below:

{\scriptsize
\begin{verbatim}
...
<a href="http://www.readwriteweb.com/cloud/2010/10/latest-ubuntu-1010-emphasizes.php" 
  id="p-1">Latest <b>Ubuntu</b>10.10 Emphasizes the Cloud - ReadWriteCloud</a> 
  <table border="0" cellpadding="0" cellspacing="0"> 
    <tbody>
...
\end{verbatim}
}

As we can see in the above excerpt, all the information we need for title and address is included  in the first $<$a ... $/$a$>$ line. To parse the information, in line 10, the WebHarvest’s XPath engine is called with the XPath expression //a[contains(@id,``p-'')] as argument which returns the title
of the result. Similarly, in lines 14-18, we acquire the abstract of the result.

The advantage of using scraping tools to wrap Web data sources is that they simplify the interfacing 
with the data sources, since no hardcoded text processing code in needed.   
While technologies like Web services have become popular nowadays, scraping tools will always be necessary to get information form data sources that isn’t yet offered through some SOAP-like interface.

\noindent\textbf{Presentation and visualization}.
\texttt{FreePub} provides several facet-based visualization and presentation options to manipulate the resulting list of documents and their support material. The results may be
organized by date, forum, author, or using any regular expressions that involves any of the above fields.
Note that any time during a creativity cycle, the user may
import any of the result (i.e., document, support material, etc) into the mindmap.   


\section{Semantic query expansion}
\label{sec:sqe}

In \texttt{FreePub}, query formulation is performed by extracting keywords from
mindmap elements. The whole task is coordinated by a {\emph semantic query expansion} mechanism. The key point is that keywords are not extracted only from user-selected mindmap elements, but also from their \emph{semantic neighbourhood}. 

Initially, the semantic neighbourhood
is decided automatically by the system, and includes important elements which are connected with the 
selected elements in the mi\-nd\-map. The user may refine the neighbourhood by marking/unmarking 
mindmap elements. 

\begin{figure}[!ht]
{\center
\fbox{\includegraphics[width=12cm]{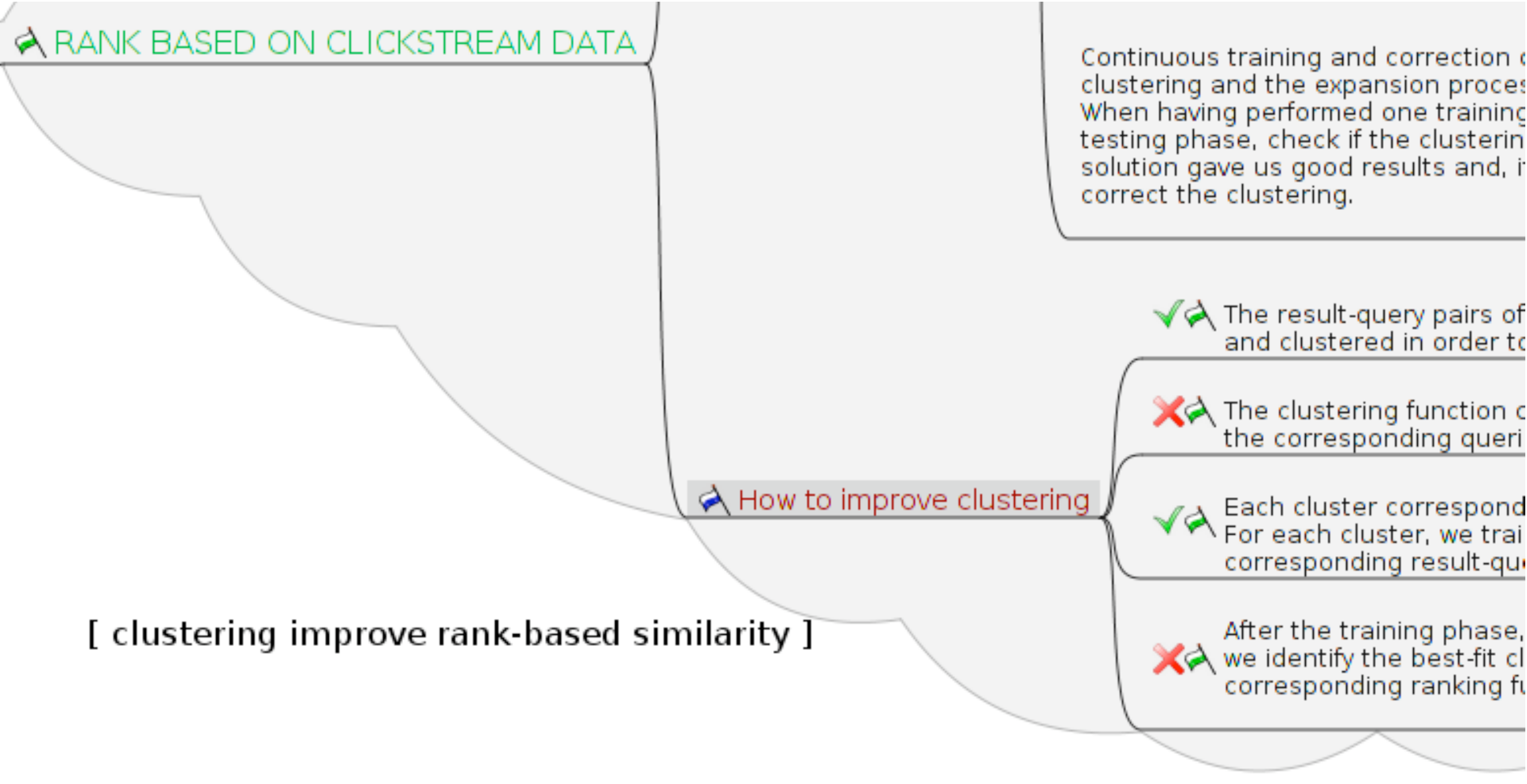}}
\caption{An example of semantic query expansion.}
\label{fig:semantic-expansion-example}
}
\end{figure}

\texttt{FreePub} employs a \emph{term ranking scheme} to determine the top-K important
terms (i.e., keywords) in the semantic neighbourhood of user-selected mindmap elements. 
These terms are used to expand the initial keyword query. Term importance is decided based on a 
tf/idf-oriented weighting sche\-me~\cite{irbook}. 
Terms are ordered accoring to their importance and the top-K terms are selected to expand the 
initial query. See for example Figure \ref{fig:semantic-expansion-example}, where the user has 
selected the mindmap element ``How to improve clustering" (marked by the system using a blue flag). 
Note that the system has also marked other mindmap elements around (marked using 
a green flag).
These latter elements form the semantic neighbourhood of the selected element.
Finally, the terms considered by the system for the query expansion are ``clustering improve 
rank-based similarity''.
Next we describe in detail how we determine the query expansion terms:
\begin{enumerate}
\item All elements in the neighbourhood of user-selected elements are considered as documents
and are indexed using the Lucene IR engine~\cite{lc}. The level of neighbourhoud is user defined, e.g., level $1$ means that the neighbourhood of a selected element includes
only directly adjacent nodes.
\item To each document $d$, we assign weights $docWeight_d$ according to the type of corresponding elements. For example, a document that is formed from topic elements gets higher
weight than that formed from detail elements (see Table \ref{tab:10}).
\item Terms are cleaned (i.e., punctuation and stopwords are removed), and the
number of terms $docSize_d$ for each document $d$ is calculated.
\item For each term $t$, we compute its number $freq_t^d$ of occurences in each doc $d$ (i.e., term frequency), and the number $docFrec_t$ of documents containing term $t$.
\item Then, we compute, for each term $t$, its score $w^d_t$ for every document $d$:
$w^d_t=\frac{freq_t^d\times docFreq_t\times docWeight_d}{docSize_d}$. The final score $W_t$ for each term $t$ is the average of its scores $w^d_t$.
\item Terms are sorted according to $W_t$, and the terms with the better K scores
are used to expand the initial query. K is user-defined.
\end{enumerate}

\section{\texttt{FreePub} in use}
\label{sec:freepub}

Since there are no mindmap benchmarks, we demonstrate \texttt{FreePub} 
advantages by presenting in this section a test case of working with \texttt{FreePub}
(arranged with the research team of DIANA lab\footnote{http://microrna.gr/} at BSRC Fleming)
to collect and organize scientific material regarding the microRNA target prediction problem.

\begin{figure}[!ht]
{\center
\fbox{\includegraphics[width=12cm]{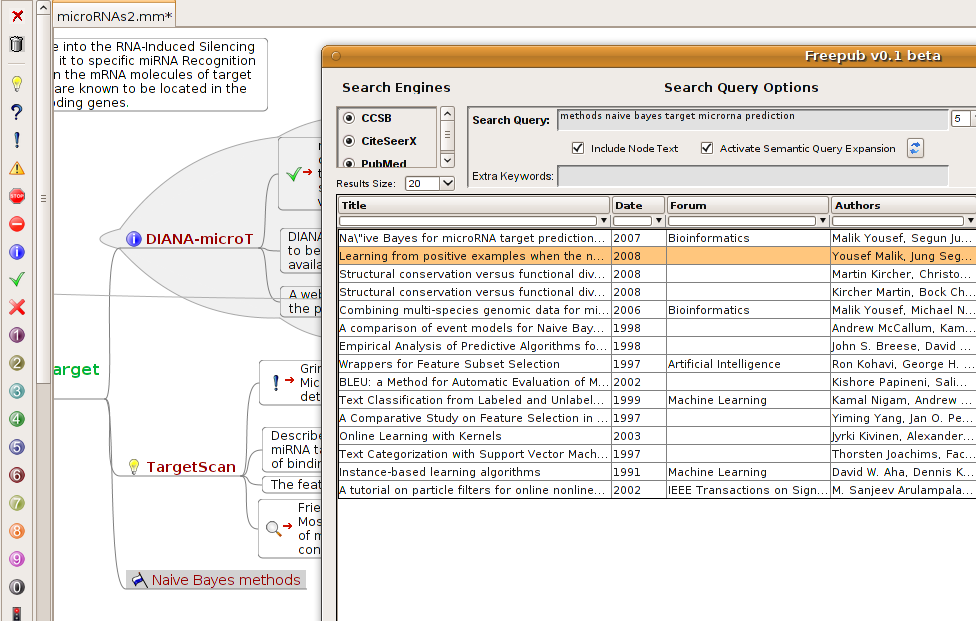}}
\caption{Searching with \texttt{FreePub}.}
\label{fig:search-example}
}
\end{figure}

\begin{figure}[!ht]
{\center
\fbox{\includegraphics[width=12cm]{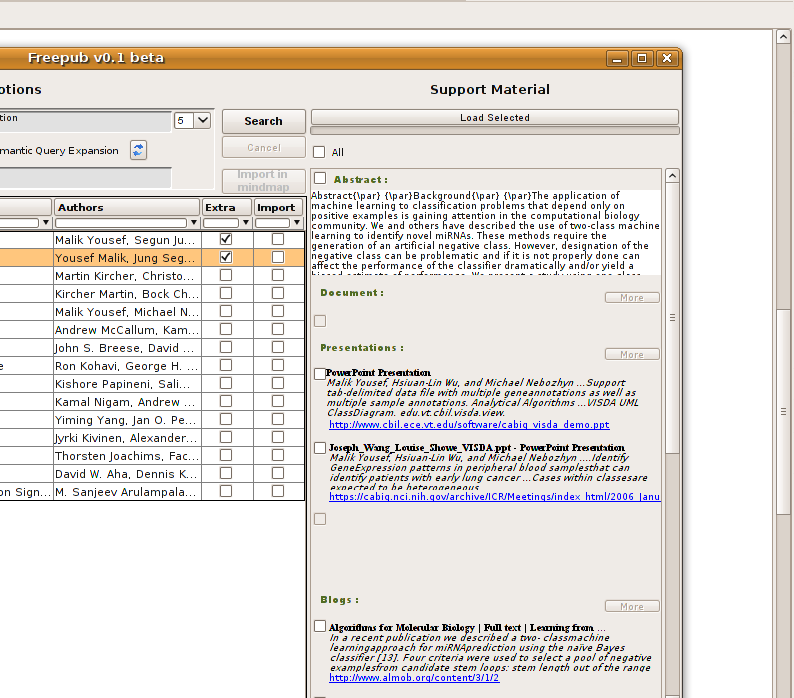}}
\caption{Searching with \texttt{FreePub}: the supporting material.}
\label{fig:support-example}
}
{\center
\fbox{\includegraphics[width=12cm]{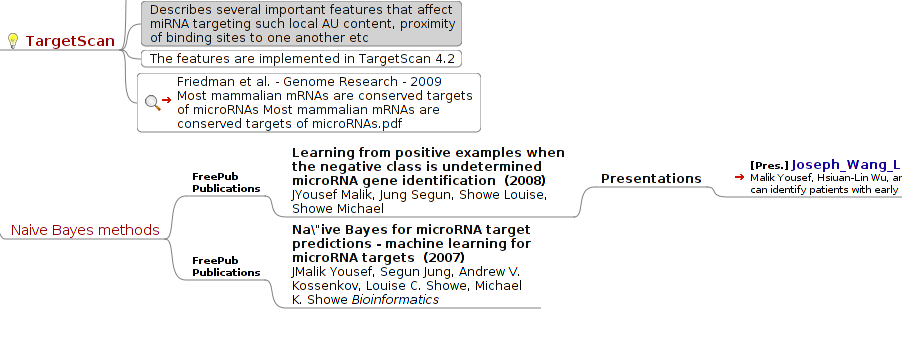}}
\caption{Importing results in mindmap.}
\label{fig:import-example}
}
\end{figure}

Next, we give some background info for microRNAs to better understand the mindmap in
Figure \ref{fig:mindmap-example}. 
microRNAs (miRNAs) are short RNA molecules
that regulate gene expression by binding directly and preferably to the 3' untranslated region 
(3'UTR) of the sequence of genes~\cite{bartel-cell04}. Each mature miRNA is 19-24
nucleotides in length, and is processed from longer 70-nucleotide stem-loop structures known as
pre-miRNAs.
Pre-miRNAs are processed to mature miRNAs in the cytoplasm by interaction with the endonuclease Dicer. Each miRNA is integrated
into the RISC (RNA induced silencing complex) complex and guides the whole
complex to the mRNA sequence of a gene, thus inhibiting translation or inducing mRNA
degradation~\cite{liu-science04}. Since their initial identification, miRNAs have been found to confer a novel layer of genetic regulation in a wide range of biological processes. MiRNAs were
first identified in 1993~\cite{lee-cell93} via classical genetic techniques in C. elegans, but it was not until 2001 that they were found to be widespread and
abundant in cells~\cite{lagos-science01}. This finding served as the primary impetus for the
development of the first computational miRNA target prediction programs. DIANA-
microT~\cite{kiriakidou-genes04} and TargetScan~\cite{lewis-cell05} were the first algorithms to
predict miRNA target genes in humans, and led to the identification of an initial set of experimentally supported mammalian targets. Such targets are now collected and reported in
TarBase~\cite{papadopoulos-nar08} which contains more than one thousand entries for human and
mouse miRNAs.

Figure \ref{fig:mindmap-example} illustrates part of a mindmap for the miRNA target prediction problem set up by the researchers.
Take for example the mindmap element \texttt{microRNA target prediction}, and its two subtopic
elements \texttt{DIANA-microT} and \texttt{TargetScan}. Both predict genes that are 
targeted by miRNAs. The former was introduce in 2004, and since then it has received 
significant improvements. Currently has been shown (using pSilac) to be the most precise program
currently available. The latter provides several important features that affect miRNA
targeting.


Generally, most target prediction programs use several features to identify putative miRNA binding
sites, such as evolutionary conservation, structural accessibility, nucleotide composition and 
others. Thus, a researcher considers that training learning functions using Naive Bayes
models might be one way to follow for miRNA target prediction. She records this as
a mindmap element, and starts the search. Figure \ref{fig:search-example} shows the resulting list of papers.
Note that \texttt{FreePub} has expanded the initial user query from ``Naive Bayes''
to ``methods naive bayes target microrna prediction'', due to its semantic query expansion
service.

The researcher selects, then, a couple of papers and a related presentation as supporting material to move to the mindmap. 
Figure \ref{fig:support-example} shows the retrieved supporting material, and Figure \ref{fig:import-example} shows the resulting mindmap.

\section{Current status and future work}
\label{sec:con}
In this work, we presented \texttt{FreePub}, a creativity support tool to collect and organize
scientific material using mindmaps. \texttt{FreePub} supports creativity cycles. A user starts such a cycle by setting up her domain of interest using mindmaps. 
Then, she can browse mindmaps and launch search tasks to gather relevant publications from several data sources. \texttt{FreePub}, besides publications,  identifies helpful supporting material (e.g., blog posts, presentations). All retrieved information from \texttt{FreePub} can be organized in mindmaps.
\texttt{FreePub} has been fully implemented on top of FreeMind, a popular open-source, mindmapping tool.

For future work, we first plan to set up a detailed user-based evaluation of our tool with the help
of a large number of scientists, and record their feedback after performing creativity cycles using \texttt{FreePub}. We also plan to develop several new services: (a) tagging facilities, (b) retrieval facilities for support material like, e.g., survey papers, highly-impact papers, etc., and (c) visual, easy-to-use scrapping facilities based on user query-by-example input in order to wrap data sources.
Moreover, we will work on improving the semantic query expansion method. Finally, we will exploit public services like
Mendeley and
CiteULike\footnote{http://www.mendeley.com/, http://www.citeulike.org/} to
evaluate the impact of retrieved publications, and the relations between them.

\bibliographystyle{plain} 
\bibliography{mm} 

\end{document}